\documentclass[prd,preprintnumbers,12pt]{revtex4}
\pdfoutput=1
\usepackage{epsfig}
\usepackage{graphicx}
\usepackage{bm}

\newcommand{\be}{\begin{equation}}
\newcommand{\dd}{\displaystyle}
\newcommand{\ee}{\end{equation}}
\newcommand{\bea}{\begin{eqnarray}}
\newcommand{\eea}{\end{eqnarray}}

\newcommand{\nn}{\nonumber}
\newcommand{\de}{\partial}
\def\nn{\nonumber}
\def\de{\partial}

 \def\slash#1{\setbox0=\hbox{$#1$}#1\hskip-\wd0\dimen0=5pt\advance
       \dimen0 by-\ht0\advance\dimen0 by\dp0\lower0.5\dimen0\hbox
         to\wd0{\hss\sl/\/\hss}}
\def\be{\begin{equation}}
\def\dd{\displaystyle}
\def\ee{\end{equation}}

\def\bea{\begin{eqnarray}}
\def\eea{\end{eqnarray}}
\def\tG{\tilde G}
\def\7{\tilde}
\def\8{\hat}

 \def\slash#1{\setbox0=\hbox{$#1$}#1\hskip-\wd0\dimen0=5pt\advance
       \dimen0 by-\ht0\advance\dimen0 by\dp0\lower0.5\dimen0\hbox
         to\wd0{\hss\sl/\/\hss}}

\def\tx{\tilde x}
\def\txi{\tilde\xi}

\def\tZ{\tilde Z}
\def\tc{\tilde c}

\def\tP{\tilde P}
\def\tB{\tilde B}
\def\tc{\tilde c}
\def\tx{\tilde x}

\def\txi{\tilde\xi}

\def\tZ{\tilde Z}
\def\tc{\tilde c}

\usepackage{color}

\begin{document}

%\subheader{\hfill {\rm ICCUB-18-011}}
%\vskip0.1cm\hfill{\bf
%DFF-412/03/04\,\,}
%\preprint{{\bf ICCUB-10-174}} 
%\preprint{{\bf UB-ECM-PF 10/42}}
{\hfill {\rm ICCUB-18-024}}
\title{VSUSY models with Carroll or Galilei invariance}

\author{Andrea Barducci}\email{barducci@fi.infn.it} \affiliation{Department of Physics, University of Florence and INFN Via G. Sansone 1, 50019 Sesto Fiorentino (FI), Italy}
\author{Roberto Casalbuoni}\email{casalbuoni@fi.infn.it}
\affiliation{Department of Physics, University of Florence and INFN Via G. Sansone 1, 50019 Sesto Fiorentino (FI), Italy}
\author{Joaquim Gomis}\email{joaquim.gomis@ub.edu}\affiliation{Departament de F\'isica Qu\`antica i Astrof\'isica\\ and
Institut de Ci\`encies del Cosmos (ICCUB), Universitat de Barcelona\\ Mart\'i i Franqu\`es , ES-08028 Barcelona, Spain
}
\date{\today}


\begin{abstract}
{The general method introduced in a previous paper
% {\color{green} [1]}
 to build up a class of models invariant under generalization of Carroll and Galilei algebra
is extended to systems including a set of Grassmann variables describing the spin degree of freedom. 

The models described here are based on a relativistic supersymmetric algebra with vector and scalar generators (VSUSY) . Therefore, in order to obtain dynamical systems consistent with Carroll or Galilei, we will study the contractions of the anticommuting generators compatible with the Poincar\'e contractions.%}
}
\end{abstract}
%\pacs{12.38.-t, 26.60.+c, 74.20.-z, 74.20.Fg, 97.60.Gb}
\maketitle

\section{Introduction}

In a previous paper \cite{Barducci:2018wuj} we have introduced a general strategy to build up a  class of models invariant under generalizations of Carroll and Galilei algebra with zero central charge
\cite{Levy-Leblond,Sou,Duval:2009vt,Duval:2014uoa,Bergshoeff:2014jla,Cardona:2016ytk,Batlle:2016iel,Gomis:2016zur,Batlle:2017cfa}
%\cite{Gomis:2000bd,Gomis:2005pg,Brugues:2004an,Brugues:2006yd,Bergshoeff:2014jla,Batlle:2016iel,Cardona:2016ytk}. 
%{\color{blue} 
A bonus of this approach is that it allows a description in configuration space, whereas most  of the models invariant under Carroll or Galilei group present in the literature are described by an action  in phase space. Furthermore, although the construction of these nonrelativistic actions starts from a relativistic formulation, the method in  \cite{Barducci:2018wuj} does not require any limiting procedure and redefinition of the parameters.%}

Aim of this work is to apply this method to systems including a set of Grassmann variables describing the spin  degrees of freedom, exhibiting a Carroll or a Galilei symmetry. The system we have in mind is one with a SUSY symmetry described by vector-like 
 and scalar anticommuting generators (VSUSY). This algebra, which is a relativistic one, was introduced in \cite{Barducci:1976qu}, %alter stuided in detail in 
   in order to get a "pseudoclassical" description of the Dirac equation.
 %{\color{blue}
  See also \cite{Brink:1976sz,Brink:1976uf,Berezin:1976eg}, where these pseudoclassical models have the same set of Grassmann variables and lead to the Dirac equation without rigid supersymmetry.%}
 
 Of course, the models we would like to construct here should exhibit Carroll or Galilei supersymmetry
 counterpart.
To this end, we would need to construct specific contractions of  VSUSY
 algebra. Before doing that, we would need to review the general strategy of  \cite{Barducci:2018wuj}.

 The method consists in starting from a space-time in $D+1$ dimensions and partitioning it in two parts, the first minkowskian and the second euclidean. Then a Carroll 
 or Galilei invariant model can be obtained by introducing a Minkowski invariant action,
  or Euclidean invariant action respectively,
    in one  of the partitions  of the space-time (the minkowskian for Carroll or  the euclidean  for Galilei)
    and,   in the complementary partition,
     a system of lagrange multipliers transforming in an appropriate way  under the euclidean  or the Lorentz group 
 such that to confine the system to a region of the space-time.
 This system is such to compensate the variations induced by the Carroll 
  or Galilei
 boosts of the action in the  appropriate subspace. A simple way to get this result, is to start with an action Poincar\'e invariant in the total space, say
$S$, and defining the action in the Minkowski  or euclidean subspace as
\be
(S)_{\rm Carroll}=(S)|_{x^a\equiv 0}, 
\ee
or
 \be
 (S)_{\rm Galilei}=(S)|_{x^\alpha\equiv 0}, 
 \ee
where the $x^a$'s and ${x^\alpha}$'s are the coordinates of the euclidean subspace  where the system is confined respectively for the Carroll and Galilei case.

%and adding to it terms in the euclidean subspace obtained by looking at the variation of  $(S)_{\rm Carroll}$ under the Carroll boosts and such to make the total action Carroll invariant.
% 
% The same procedure can be used for the Galilei case. This time using a  lagrangian defined in the euclidean sector and enlarging it with a system of lagrange multipliers living in the first part of the space-time
% {\color{magenta}.not complete clear tome the previous sentence. We repit ourselves in this concept.. I will be shorther} Again, a simple way is to start with the euclidean version, $S_E$, of an action Poincar\'e invariant in the total space, and define the action in the euclidean subspace as
% \be
% (S)_{\rm Galilei}=(S_E)|_{x^\alpha\equiv 0}, 
% \ee
% where the $x^\alpha$'s are the coordinates of the Minkowski subspace. The total action is obtained following the same lines outlined for the Carroll case. 

It is important to notice that this procedure allows us to obtain the action, for dynamical systems invariant under Carroll or Galilei group without central charges, directly from actions in configurations space invariant under Poincar\'e without use of a limiting procedure.
 %  whereas in most of the cases examined in ref. \cite{Barducci:2018wuj}, one needs to start from phase space actions, invariant inder Poincar\'e. in order to perform the limiting provrdutr to get Carroll or Galilei invariant action. 
  %This can be easily understood in our approach. As 
As we have said one needs to introduce a set of lagrange multipliers in one of the two parts
of  the lagrangian  in which we divide the space-time. These multipliers have not a clear interpretation in configuration space.
%and certainly they can br obtained fromn the limit of an action invariant under Poincar\'e. 
However, in general,  these lagrange multipliers have a natural interpretation as momenta in the 
phase space canonical action.

 The paper is organized as follows: in Section 2 we review the $k$-contractions  \cite{Gomis:2000bd, Danielsson:2000gi, Brugues:2004an, Gomis:2005pg, Brugues:2006yd}
 {\cite{Batlle:2016iel,Cardona:2016ytk,Gomis:2016zur,Batlle:2017cfa}
 used in ref. \cite{Barducci:2018wuj}. In Sections 3.1-3.4, we construct the $k$-contractions of VSUSY algebra for the Carroll type  and the Galilei type, both in the case of the abstract algebra and in the case of the realization of the algebra in the configuration space. In Section 4, following the procedure illustrated in \cite{Barducci:2018wuj} we construct the action and discuss the case for a VSUSY Carroll particle. In particular we examine its $\kappa$ invariance and its quantization respectively in Sections 4.1 and 4.2. The same is done for a VSUSY Galilei particle in Sections 5, 5.1 and 5.2.
In Section 6 we draw our conclusions. In the Appendix we show how the actions, for the VSUSY Carroll and Galilei particle, can be derived by performing the standard Carroll and non-relativistic limiting procedures in the Poincar\'e invariant phase space action.

\section{Poincar\'e contractions}

In order to make explicit the partition of the space-time in $D+1$ dimensions we will introduce the following set of coordinates
\bea
&&\mu,\nu=0,1,...,D,~~~\eta_{\mu\nu}=(-;+,\cdots,+), \nn\\
&&\alpha,\beta =0,1,\cdots,k-1,~~~\eta_{\alpha\beta}=(-;+,\cdots,+),\nn\\
&&a,b =k,\cdots,D,~~~\eta_{ab}=(+,+,\cdots,+),
\eea
The Poincar\'e algebra in the total space is given by
 \bea
 \left[M_{\mu\nu},M_{\rho\sigma}\right]&=&-i(\eta_{\mu\rho}M_{\nu\sigma}+\eta_{\nu\sigma}M_{\mu\rho}
 -\eta_{\mu\sigma}M_{\nu\rho}
  -\eta_{\nu\rho}M_{\mu\sigma}),
\nonumber \\
  \left[M_{\mu\nu},P_{\rho}\right]&=&-i(\eta_{\mu\rho}P_{\nu}-\eta_{\nu\rho}P_{\mu}),
\nn\\
  \left[P_{\mu},P_{\nu}\right]&=& 0 \ ,\label{eq:1.1}
\eea
Then, consider  the following two subgroups of $ISO(1,D)$: the Poincar\'e subgroup in $k$ dimensions, $ISO(1,k-1)$ and the euclidean group of roto-translations in $D+1-k$ dimensions, generated respectively by
\be
ISO(1,k-1):~~~M_{\alpha\beta},~~P_\alpha,~~~\alpha,\beta =0,1,\cdots,k-1,
\ee
\be
ISO(D+1-k):~~~M_{ab},~~P_a,~~~a,b =k,\cdots,D.
\ee
In these notations the generators of $ISO(1,D)$ are
\be
ISO(1,D):~~~M_{\alpha\beta},~~~M_{ab},~~P_\alpha,~~~P_a,~~~M_{\alpha b}\equiv B_{\alpha b}.
\ee
 Note that the generators $ B_{\alpha b}$ connect the two subalgebras.

In \cite{Barducci:2018wuj} we have considered  two types of contractions, both at the level of the Poincar\'e algebra and at the level of 
 the invariant vector fields
. These contractions generalize the Carroll \cite{Levy-Leblond, Bergshoeff:2014jla, Duval:2014uoa, Cardona:2016ytk}  and the Galilei algebras 
\cite{LL} \cite{Gomis:2000bd, Danielsson:2000gi, Brugues:2004an, Gomis:2005pg, Brugues:2006yd} \cite{Batlle:2016iel,Gomis:2016zur,Batlle:2017cfa}

At the {\color{blue} Lie }algebra level the contractions are made on the momenta and on the boosts; in the Carroll case:
\be
\tilde P_\alpha={\dd \frac 1\omega} P_\alpha,~~~
\tilde B_{\alpha a}= {\dd \frac 1\omega} B_{\alpha a}\label{eq:2.6}
\ee
and then taking the limit $\omega\to\infty$. In the Galilei case the contractions are given by
\bea
\tilde P_a ={\dd \frac 1\omega}P_a,,~~~
\tilde B_{\alpha a}= {\dd \frac 1\omega} B_{\alpha a}.\label{eq:2.7}
\eea

 The commutation relations resulting from the contraction process are:\\\\

 \noindent {\bf Carroll-type}
\be
\left(\begin{array}{c|ccc}
 & B_{\beta b} & P_\beta & P_b \\
\hline
B_{\alpha a} & 0 & 0 & i\eta_{ab}P_\alpha \\
P_\alpha & 0 & 0 & 0 \\
P_a & -i\eta_{ab}P_\beta & 0 & 0
\end{array}\right),
 \ee
 \\\\
\noindent
{\bf Galilei-type}
\be
\left(\begin{array}{c|ccc}
 & B_{\beta b} & P_\beta & P_b \\
\hline
B_{\alpha a} & 0 & -i\eta_{\alpha\beta}P_a &  \\
P_\alpha & i\eta_{\alpha\beta}P_a & 0 & 0 \\
P_a & 0 & 0 & 0
\end{array}\right).
 \ee

\section{VSUSY contractions}

Let us now consider the VSUSY algebra \cite{Barducci:1976qu, Casalbuoni:2008iy} which is defined in terms of  the generators $G_\mu$ and $G_5$ and of their anticommutation relations
\be
[G_\mu,G_\nu]_+ =\eta_{\mu\nu}Z,~~~[G_\mu,G_5]_+=-P_\mu,~~~[G_5,G_5]_+= Z_5
\ee
 By using a decomposition analogous to the one used in the Poincar\'e
 case
\be
G_\mu=(G_\alpha,G_a),~~~\alpha =0,1,\cdots,k-1,~~~a =k,\cdots,D
\ee
The anticommutation relations become
\be
[G_\alpha,G_\beta]_+=\eta_{\alpha\beta}Z,~~~[G_a,G_b]_+=\eta_{ab}Z,~~~[G_5,G_5]_+=Z_5\label{eq:1.3}
\ee
\be
[G_\alpha,G_a]_+=0,~~~[G_\alpha, G_5]_+=-P_\alpha,~~~[G_a, G_5]_+=-P_a\label{eq:1.4}
\ee
Both $G_\alpha$ and $G_a$ behave as vectors under the two groups generated by $M_{\alpha\beta}$ and $M_{ab}$. Therefore the corresponding commutation relations are invariant under any rescaling of $G_\alpha$ and $G_a$. This is not the same for the Lorentz boosts with commutation relations
\be
[B_{\alpha a}, G_\beta]=+i\eta_{\alpha\beta}G_a,~~~[B_{\alpha a}, G_b]=-i\eta_{ab}G_\alpha\label{eq:1.5}\ee

\subsection{$k$-contractions of Carroll type}

In the Poincar\'e case we have defined the contraction by a rescaling of part of the momenta and for the boosts, according to the eq. (\ref{eq:2.6}) for the Carroll case and eq. (\ref{eq:2.7}) for the Galilei one,
leaving unchanged all the other generators. By doing so we have not changed the two  subalgebras 
$ISO(1,k-1)$ and $ISO(D+1-k)$. In the case of VSUSY we would like to make use of  an analogous strategy. In particular , for the Carroll case,
we would like to maintain the VSUSY invariance in the Minkowski sector. Therefore
we will study $k$-contractions  of  Carroll type, preserving the main anti-commutation relation in this sector:
\be
[G_\alpha, G_5]_+=-P_\alpha\label{eq:1.7}
\ee
To this end we will define the contracted generators as
\be
\tilde G_\alpha =\frac 1 {\omega^r} G_\alpha,~~~ \tilde G_a=\frac 1{\omega^t}G_a,~~~\tilde G_5 =\frac 1{\omega^{1-r}}G_5\label{eq:1.8}
\ee
With this choice we have
\be
[ \tG_\alpha,\tG_\beta]_+=\eta_{\alpha\beta}\frac Z{\omega^{2r}},~~~[\tG_a,\tG_b]_+=\eta_{ab}\frac Z{\omega^{2t}},~~~[\tG_5,\tG_5]_+=\frac {Z_5}{\omega^{2-2r}}
\ee
\be
[\tG_\alpha, \tG_5]_+=-\tP_\alpha,~~~[\tG_a, \tG_5]_+=-\frac 1{\omega^{1-r+t}}\tP_a\label{eq:1.10}
\ee
\be
[\tB_{\alpha a}, \tG_\beta]=-\frac i{\omega^{1+r-t}}\eta_{\alpha\beta}\tG_a,~~~[\tB_{\alpha a}, \tG_b]=+\frac i{\omega^{1-r+t}}\eta_{ab}\tG_\alpha\label{eq:1.11}
\ee
where, for the moment being we have not done any choice regarding the possible contractions of the central charges. 
Asking to maintain the complete VSUSY algebra in the Minkowski subspace, we should scale both $Z$ and $Z_5$
\be
\tilde Z=\frac Z{\omega^{2r}},~~~ \tZ_5=\frac {Z_5}{\omega^{2-2r}}
\ee 
Then, we have
\be
[ \tG_\alpha,\tG_\beta]_+=\eta_{\alpha\beta} \tZ,~~~
[\tG_a,\tG_b]_+=\eta_{ab}\frac {\tZ}{\omega^{2(t-r)}},~~~[\tG_5,\tG_5]_+= {\tZ_5}\label{eq:3.12}
\ee
\be
[\tG_\alpha, \tG_5]_+=-\tP_\alpha,~~~[\tG_a, \tG_5]_+=-\frac 1{\omega^{1-r+t}}\tP_a\label{eq:1.14}
\ee
\be
[\tB_{\alpha a}, \tG_\beta]=-\frac i{\omega^{1+r-t}}\eta_{\alpha\beta}\tG_a,~~~[\tB_{\alpha a}, \tG_b]=+\frac i{\omega^{1-r+t}}\eta_{ab}\tG_\alpha\label{eq:3.15}
\ee
Now, all these relations depend only on the difference $r-t$. To have a finite result, it is enough to   require $t-r\ge 0$ (as it follows from (\ref{eq:3.12})) and, from  (\ref{eq:3.15}) , $1+r-t\ge 0$.
 %Then, $1+r-t\ge 0$. Therefore, in order to satisfy $1-r+t\ge 0$,  assuming integer values for this difference, 
There are only two possibilities to verify these conditions:
\\\\
1) $t-r=0$,
the relations become
\be
[ \tG_\alpha,\tG_\beta]_+=\eta_{\alpha\beta} \tZ,~~~
[\tG_a,\tG_b]_+=\eta_{ab}{\tZ},~~~[\tG_5,\tG_5]_+= {\tZ_5}
\ee
\be
[\tG_\alpha, \tG_5]_+=-\tP_\alpha,~~~[\tG_a, \tG_5]_+=0\label{eq:1.14}
\ee
\be
[\tB_{\alpha a}, \tG_\beta]=0,~~~[\tB_{\alpha a}, \tG_b]=0\label{eq:1.15}
\ee
Notice that the $G_a$'s form a Clifford algebra (after renormalization of the generators) which commute or anticommute with all other generators of the previous list. The algebra is $VSUSY\otimes Clifford$, with VSUSY in $k$-dimensions and the Clifford in $D+1-k$ dimensions.\\\\
\noindent
2) $t-r =1$, we get
\be
[ \tG_\alpha,\tG_\beta]_+=\eta_{\alpha\beta} \tZ,~~~
[\tG_a,\tG_b] _+=0,~~~[\tG_5,\tG_5]_+= {\tZ_5}
\ee
\be
[\tG_\alpha, \tG_5]_+=-\tP_\alpha,~~~[\tG_a, \tG_5]_+=0\label{eq:1.14}
\ee
\be
[\tB_{\alpha a}, \tG_\beta]=-i\eta_{\alpha\beta}\tG_a,~~~[\tB_{\alpha a}, \tG_b]=0\label{eq:1.15}
\ee
Here the $\tG_a$'s span a Clifford algebra with zero central charge, or a Grasmmann algebra.
In the  case $r=t$, corresponding to the model discussed in the following, the boosts connect only the space-time variables among the two sectors.

\subsection{$k$-contractions of Carroll type in configuration space}

The scaling of the algebra of the  generators, implies the following scaling in the configuration space  realization
\be
\tx^\alpha =\omega x^\alpha,~~~ \tx^a=x^a,~~~\txi^\alpha=\omega^r\xi^\alpha,~~~\txi^a=\omega^t\xi^a,~~~\txi^5=\omega^{1-r}\xi^5,~~~\tc=\omega^{2r} c,~~~\tc_5=\omega^{2-2r} c_5\label{eq:1.23}
\ee
We will make use of the vector field realization of the VSUSY algebra, as given
 %in the paper
  \cite{Casalbuoni:2008iy}:
\be 
G_\mu=-i\left(\frac\de{\de\xi^\mu} +i\xi^5\frac\de{\de x^\mu}-\frac i 2 \xi_\mu\frac\de{\de c}\right),~~~
G_5=-i\frac\de{\de\xi^5} -\frac 12 \xi^5\frac\de{\de c_5}\label{eq:1.24}
\ee
with the generators of the Poincar\'e  group in the $D+1$ dimensional space:
\be
M_{\mu\nu} =-i\left(x_\mu\frac\de{\de x^\nu}-x_\nu\frac\de{\de x^\mu}\right) -i
\left(\xi_\mu\frac\de{\de \xi^\nu}-\xi_\nu\frac\de{\de \xi^\mu}\right),~~~P_\mu=-i\frac\de{\de x^\mu}\label{eq:1.25}\ee
 Notice that the derivatives with respect to the Grassmann variables are left-handed derivatives.

Performing the scaling on the vector fields  according to eqs. (\ref{eq:1.8}), (\ref{eq:2.6}) and (\ref{eq:1.23}) we get for the two cases
\noindent
1) $t-r=0$,
\be
\tG_\alpha= -i\left(\frac\de{\de\txi^\alpha} +i\txi^5\frac\de{\de \tx^\alpha}-\frac i 2 \txi_\alpha\frac\de{\de \tc}\right)
\ee
\be
\tG_a= -i\left(\frac\de{\de\txi^a}-\frac i 2 \txi_a\frac\de{\de \tc}\right)\ee
\be
\tG_5=-i\frac\de{\de\txi^5} -\frac 12 \txi^5\frac\de{\de \tc_5}
\ee
\be
\tB_{\alpha a}=i\tx_a\frac\de{\de\tx^\alpha}
\ee
Implying the following variations for the coordinates (omitting the tilde)
\be
i\epsilon^\alpha G_\alpha:~~~\delta\xi^\alpha =\epsilon^\alpha,~~~\delta x^\alpha=i\epsilon^\alpha\xi^5,~~~\delta c= -\frac i2 \epsilon^\alpha \xi_\alpha\ee
\be
i\epsilon^a G_a:~~~\delta\xi^a =\epsilon^a,~~~\delta x^a=0,~~~\delta c= -\frac i2 \epsilon^a\xi_a
\ee
\be 
i\epsilon^5 G_5:~~~\delta \xi^5=\epsilon^5,~~~\delta c_5=-\frac i 2\epsilon^5\xi^5
\ee
\be
iv^{\alpha a}B_{\alpha a}:~~~\delta x^\alpha = -v^{\alpha a}x_a
\ee

\vskip1cm\noindent
2) $t-r =1$, we get
\be
\tG_\alpha= -i\left(\frac\de{\de\txi^\alpha} +i\txi^5\frac\de{\de \tx^\alpha}-\frac i 2 \txi_\alpha\frac\de{\de \tc}\right)
\ee
\be
\tG_a= -i\frac\de{\de\txi^a}\ee
\be
\tG_5=-i\frac\de{\de\txi^5} -\frac 12 \txi^5\frac\de{\de \tc_5}
\ee
\be
\tB_{\alpha a}=i\tx_a\frac\de{\de\tx^\alpha}-i
\txi_\alpha\frac\de{\de \txi^a}
\ee

Implying the following variations for the coordinates (omitting the tilde)
\be
i\epsilon^\alpha G_\alpha:~~~\delta\xi^\alpha =\epsilon^\alpha,~~~\delta x^\alpha=i\epsilon^\alpha\xi^5,~~~\delta c= -\frac i2 \epsilon^\alpha \xi_\alpha\ee
\be
i\epsilon^a G_a:~~~\delta\xi^a =\epsilon^a,~~~\delta x^a=0,~~~\delta c= 0
\ee
\be 
i\epsilon^5 G_5:~~~\delta \xi^5=\epsilon^5,~~~\delta c_5=-\frac i 2\epsilon^5\xi^5
\ee
\be
iv^{\alpha a}B_{\alpha a}:~~~\delta x^\alpha = -v^{\alpha a}x_a,~~~\delta \xi^a=v^{\alpha a}\xi_\alpha\label{eq:3.45}
\ee
 Note that the difference between the two types of transformations is the action 
of the boosts .  In the actual case, the boosts depend also on the Grassmann variables, $\xi^\alpha$ amd $\xi^a$,  Furtermore  the VSUSY algebra decomposes in the direct product of a VSUSY in the Minkowski subspace times a Grassmann algebra in the euclidean part.

\subsection{$k$-contractions of Galilei type}

In this case the contraction at the level of the Poincar\'e group 
is defined in eq.(\ref{eq:2.7}). This time we want to maintain the VSUSY invariance in the euclidean subsector. Therefore
we will study the $k$-contractions  of  Galilei  type, preserving the main anti-commutation relation:
\be
[G_a, G_5]_+=-P_a\label{eq:1.7}
\ee
To this end we will define the contracted generators as
\be
\tilde G_\alpha =\frac 1 {\omega^r} G_\alpha,~~~ \tilde G_a=\frac 1{\omega^t}G_a,~~~\tilde G_5 =\frac 1{\omega^{1-t}}G_5\label{eq:1.3.8}
\ee
With this choice we have
\be
[ \tG_\alpha,\tG_\beta]_+=\eta_{\alpha\beta}\frac Z{\omega^{2r}},~~~[\tG_a,\tG_b]_+=\eta_{ab}\frac Z{\omega^{2t}},~~~[\tG_5,\tG_5]_+=\frac {Z_5}{\omega^{2-2t}}
\ee
\be
[\tG_\alpha, \tG_5]_+=-\frac 1{\omega^{1+r-t}}\tP_\alpha,~~~[\tG_a, \tG_5]_+=-\tP_a\label{eq:1.3.10}
\ee
\be
[\tB_{\alpha a}, \tG_\beta]=-\frac i{\omega^{1+r-t}}\eta_{\alpha\beta}\tG_a,~~~[\tB_{\alpha a}, \tG_b]=+\frac i{\omega^{1-r+t}}\eta_{ab}\tG_\alpha\label{eq:1.3.11}
\ee 
Asking to maintain the complete VSUSY algebra in the euclidean sector, we need to scale both $Z$ and $Z_5$
\be
\tilde Z=\frac Z{\omega^{2t}},~~~ \tZ_5=\frac {Z_5}{\omega^{2-2t}}\label{eq:1.45}
\ee 
Then we have
\be
[ \tG_\alpha,\tG_\beta]_+=\eta_{\alpha\beta} \frac {\tZ}{\omega^{2(r-t)}},~~~
[\tG_a,\tG_b]_+=\eta_{ab}\tZ,~~~[\tG_5,\tG_5]_+= {\tZ_5}
\ee
\be
[\tG_\alpha, \tG_5]_+=-\frac 1{\omega^{1+r-t}}\tP_\alpha,~~~[\tG_a, \tG_5]_+=-\tP_a\label{eq:1.3.14}
\ee
\be
[\tB_{\alpha a}, \tG_\beta]=-\frac i{\omega^{1+r-t}}\eta_{\alpha\beta}\tG_a,~~~[\tB_{\alpha a}, \tG_b]=+\frac i{\omega^{1-r+t}}\eta_{ab}\tG_\alpha\label{eq:1.3.15}
\ee
Reasoning as in the Carroll case, we have only two possibilities:\\\\
1) $r-t=0$,
the relations become
\be
[ \tG_\alpha,\tG_\beta]_+=\eta_{\alpha\beta} \tZ,~~~
[\tG_a,\tG_b]_+=\eta_{ab}{\tZ},~~~[\tG_5,\tG_5]_+= {\tZ_5}
\ee
\be
[\tG_\alpha, \tG_5]_+=0,~~~[\tG_a, \tG_5]_+=-P_a\label{eq:1.3.16}
\ee
\be
[\tB_{\alpha a}, \tG_\beta]=0,~~~[\tB_{\alpha a}, \tG_b]=0\label{eq:1.3.17}
\ee
Notice that the $\tG_\alpha$'s form a Clifford algebra (after renormalization of the generators). The algebra is $VSUSY\otimes{\rm  Clifford}$, with VSUSY in $D+1-k$-dimensions and the Clifford in $k$ dimensions.\\\\

\noindent
2) $r-t =1$, we get
\be
[ \tG_\alpha,\tG_\beta]_+=0,~~~
[\tG_a,\tG_b] _+=\eta_{ab}\tZ,~~~[\tG_5,\tG_5]_+= {\tZ_5}
\ee
\be
[\tG_\alpha, \tG_5]_+=0,~~~[\tG_a, \tG_5]_+=-P_a\label{eq:1.14}
\ee
\be
[\tB_{\alpha a}, \tG_\beta]=0,~~~[\tB_{\alpha a}, \tG_b]=i\eta_{ab}\tG_\alpha\label{eq:1.15}
\ee
Here the $\tG_\alpha$'s span a Clifford algebra with zero central charge, or a Grasmmann algebra. Then we can make considerations analogous to the ones made in the Carroll case. That is:
in the first case (corresponding to the model discussed in the following), the boosts connect only the space-time variables among the two sectors.

\subsection{$k$-contractions of Galilei type in configuration space}\label{sec:3.4}

The scaling of the algebra of the  generators, implies the following scaling in the configuration space
\be
\tx^a =\omega x^a,~~~\txi^\alpha=\omega^r\xi^\alpha,~~~\txi^a=\omega^t\xi^a, \txi^5=\omega^{1-t}\xi^5,~~~\tc=\omega^{2t} c,~~~\tc_5=\omega^{2-2t} c_5\label{eq:1.23}
\ee
Recalling the espressions (\ref{eq:1.24}) and (\ref{eq:1.25})
we perform  the scaling on the expression of the generators according to eqs. (\ref{eq:1.3.8}) , (\ref{eq:1.45}) and (\ref{eq:2.7}) . The result is\\\\
\noindent
1) $r-t=0$,
\be
\tG_\alpha= -i\left(\frac\de{\de\txi^\alpha}-\frac i 2 \txi_\alpha\frac\de{\de \tc}\right)
\ee
\be
\tG_a= -i\left(\frac\de{\de\txi^a}+i\txi^5\frac{\de}{\de\tilde x^a}-\frac i 2 \txi_a\frac\de{\de \tc}\right)\ee
\be
\tG_5=-i\frac\de{\de\txi^5} -\frac 12 \txi^5\frac\de{\de \tc_5}
\ee
\be
\tB_{\alpha a}=-i\tx_\alpha\frac\de{\de\tx^a}
\ee
Implying the following variations for the coordinates (omitting the tilde)
\be
i\epsilon^\alpha G_\alpha:~~~\delta\xi^\alpha =\epsilon^\alpha,~~~\delta x^\alpha=0,~~~\delta c= -\frac i2 \epsilon^\alpha \xi_\alpha\ee
\be
i\epsilon^a G_a:~~~\delta\xi^a =\epsilon^a,~~~\delta x^a=i\epsilon^a\xi^5,~~~\delta c= -\frac i2 \epsilon^a\xi_a
\ee
\be 
i\epsilon^5 G_5:~~~\delta \xi^5=\epsilon^5,~~~\delta c_5=-\frac i 2\epsilon^5\xi^5
\ee
\be
iv^{\alpha a}B_{\alpha a}:~~~\delta x^a = v^{\alpha a}x_\alpha\label{eq:3.72}
\ee
\vskip1cm\noindent
2) $r-t =1$, we get
\be
\tG_\alpha= -i\frac\de{\de\txi^\alpha} \ee
\be
\tG_a=  -i\left(\frac\de{\de\txi^a}+i\txi^5\frac{\de}{\de\tilde x^a}-\frac i 2 \txi_a\frac\de{\de \tc}\right)\ee
\be
\tG_5=-i\frac\de{\de\txi^5} -\frac 12 \txi^5\frac\de{\de \tc_5}
\ee
\be
\tB_{\alpha a}=-i\tx_\alpha\frac\de{\de\tx^a}-i
\txi_a\frac\de{\de \txi^\alpha}
\ee
Implying the following variations for the coordinates (omitting the tilde)
\be
i\epsilon^\alpha G_\alpha:~~~\delta\xi^\alpha =\epsilon^\alpha,~~~\delta x^\alpha=0,~~~\delta c= 0\ee
\be
i\epsilon^a G_a:~~~\delta\xi^a =\epsilon^a,~~~\delta x^a=i\epsilon^a\xi^5,~~~\delta c= -\frac i2 \epsilon^a\xi_a
\ee
\be 
i\epsilon^5 G_5:~~~\delta \xi^5=\epsilon^5,~~~\delta c_5=-\frac i 2\epsilon^5\xi^5
\ee
\be
iv^{\alpha a}B_{\alpha a}:~~~\delta x^a = v^{\alpha a}x_\alpha,~~~\delta\xi^\alpha = v^{\alpha a}\xi_a\ee

 In the second case,  the boosts depend also on the Grassmann variables $\xi^\alpha$ and $\xi^a$. Furthermore  the VSUSY algebra decomposes in the direct product of a VSUSY in the euclidean subspace times a Grassmann algebra in the Minkowski part.

\section{The VSUSY Carroll particle}

 We start with a lagrangian invariant under the VSUSY algebra on the total space-time in $D+1$ dimensions \cite{Casalbuoni:2008iy}
\be
L=-M\sqrt{-(\dot x^\mu-i\xi^\mu\dot\xi^5)^2}-\beta\left(\dot c+\frac i2\xi_\mu\dot \xi^\mu\right)-\gamma\left(\dot c_5+\frac i2\xi^5\dot\xi^5\right)\label{eq:4.0}
\ee

%We atert again from the VSUSY invariant lagrangian in $D+1$ dimensions given in eq. (\ref{eq:4.1}). 
This time we will consider a 1-contraction of Carroll type in the case $r-t=0$. The Carroll invariant lagrangian is obtained restricting 
(\ref{eq:4.0}) to the one-dimensional space spanned by $x^0$ and adding its variation under a boost times a lagrange multiplier. Since in this case the boost is operating only on $x^0$, $\delta x^0=-\vec v\cdot\vec x$ (see eq. (\ref{eq:3.45})), we get
\be
L=-M\sqrt{(\dot{ x^0}-i\xi^0\dot\xi^5)^2}-\beta\left(\dot c+\frac i 2\xi_0\dot{\xi^0}\right)-\gamma\left(\dot c_5+\frac i2\xi^5\dot\xi^5\right)+\vec\lambda\cdot \dot {\vec x}\label{eq:21}
\ee
This lagrangian  is in fact VSUSY invariant under the following rigid transformations (see \cite{Casalbuoni:2008iy})
\be
\delta x^0=i\epsilon\,\xi^5,~~~\delta\xi^0 =\epsilon,~~~ \delta\xi^5=\epsilon^5
\ee
\be
\delta c=-\frac i2 \epsilon\xi^0,~~~\delta c_5=-\frac i 2\epsilon^5\xi^5
\ee
The generators of these transformations satisfy the algebra
\be
[G_0,G_0]_+ =-Z,~~~[G_5,G_5]_+ =Z_5,~~~[G_0, G_5]_+ =-P_0
\ee
whre $Z$ and $Z_5$ are two central charges.
 The variation of   (\ref{eq:21})  under Carroll bosts is given by (remember that $\vec x$ does not transform under a Carroll boost)
\be
\delta L= p_0\delta {\dot  x}^0 +\delta\vec\lambda\cdot {\dot {\vec x}}= - p_0\vec v\cdot{\dot {\vec x}} + +\delta\vec \lambda\cdot {\dot {\vec x}}
\ee
where $p_0$  is the  
  canonical momentum associated to $\dot{ x}^0$
\be 
 p_0=\frac{\partial L}{\partial {\dot  x}^0}=-M\frac{(\dot{ x}^0-i\xi^0\dot\xi^5)}{\sqrt{(\dot{ x^0}-i\xi^0\dot\xi^5)^2}}
\ee
The lagrangian is Carroll invariant assuming
\be
\delta\vec\lambda=p_0\vec v
\ee
The other canonical momenta are
\be
\pi_0=\frac{\partial L}{\partial {\dot\xi}^0}=\frac{i\beta}2\xi_0
\ee
\be \pi^5=\frac{\partial L}{\partial {\dot\xi}^5}=\frac{i\gamma}2\xi^5 +i p_0\xi^0
\ee 
From these equations the following constraints follow
\be
\phi=p_0^{\,2}-M^2 
\ee
\be
\chi=\pi^5-\frac{i\gamma}2\xi^5-i p_0 \xi^0
\ee
\be
\chi_0=\pi_0-\frac{i\beta}2\xi_0\label{4.12}
\ee
We recall that the Poisson brackets for Grassmann variables are defined by
\be
\{\pi_0,\xi^0\}=-1
\ee
 The   constraint (\ref{4.12}) is  second class. We then define Dirac brackets in the usual way and, in particular, we find
\be
\{\xi_0,\xi_0\}^*= -\frac i\beta\ee
We are now able to evaluate the Dirac brackets of the odd constraint $\chi$ with itself. We find
\be
\{\chi,\chi\}^*=\frac i\beta(p_0^{\,2}+\beta\gamma)\ee
Therefore, if we choose 
\be
\gamma=-\beta=M
\ee 
the constraints $\phi$ and $\chi$ are first class and we expect, as in the relativistic case, a local  kappa symmetry of the lagrangian.

This model represents the  spinning generalization of the Carroll particle  model studied 
in \cite{Bergshoeff:2014jla}.
Although the two cases $t-r=0$ and $t-r=1$ correspond to two different algebras, it is easy to check that  the lagrangian (\ref{eq:21}) is invariant under also under the algebra corresponding to $r-t=1$.

\subsection{$\kappa$-invariance}
As we have seen, the model has two local symmetries, generated by the constraints $\phi$ and $\chi$. 
%if the condition $\beta\gamma=M^2$ holds, which can be satisfied by the choice $\beta=\gamma=M$. 
The local symmetry generated by the odd constraint is a "$\kappa$-symmetry". The transformations generated by $\chi$ are given by
\be
\delta x^0=i\kappa\xi^0,~~~\delta\xi^0=\kappa\frac{ p^0}M,~~~\delta\xi^5=+\kappa
\ee
from which
\bea
&\dd{\delta L=i\kappa( p_0\dot\xi^0+M\dot\xi^5)-\frac i 2\kappa p_0\dot \xi^0-\frac i2\xi_0 \frac d{d\tau}(\kappa p_0)-\frac i2 M\kappa\dot\xi^5-\frac icM\xi^5\dot k}&\nn\\
&\dd{=\frac i2\kappa p_0\dot\xi^0-\frac i2\xi^0\frac d{d\tau}(\kappa p_0)-\frac i2M\xi^5\dot\kappa+\frac i2 M\kappa\dot\xi^5=\frac i2\frac d{d\tau}[\kappa( p_0\xi^0+M\xi^5)]}&
\eea
Therefore the lagrangian (\ref{eq:21}) is quasi-invariant under this $\kappa$-transformation.

\subsection{Quantization}

To perform the canonical quantization %proceed as in the tachionic case. Precisely 
 we require 
 the following conditions for the operators corresponding to the Grassmann variables \cite{Casalbuoni:1975bj}
\be
[\hat\pi^5,\hat\pi^5]_+=[\hat\xi^5,\hat\xi^5]_+=[\hat\xi^5,\hat\xi^0]_+=[\hat\xi^0,\hat\pi^5]_+=0,~~~[\hat\pi^5,\hat\xi^5]_+=-i,~~~[\hat\xi_0,\hat\xi_0]_+=-\frac 1 M\label{eq:3.2},
\ee
%We proceed as in the tachionic case, using
 we give the following definitions
\be
\hat\pi^5=i\sqrt{\frac M2} P_1,~~~\hat\xi^5=\sqrt{\frac 2M} P_2,~~~\hat\xi_0=\frac i{\sqrt{2M}}\gamma_0 \gamma_5\ee
where we define
the matrix \cite{Freedman:2012zz}
\be
\gamma_5=(-i)^{m+1}\gamma_0\gamma_1\cdots\gamma_D\label{eq:4.22}
\ee
 where $m =(D+1)/2$, The matrx $\gamma_5$
 is hermitian and satisfies
\be
\gamma_5^2=1
\ee in any space-time of even dimension.
The matrices $P_1$ and $P_2$ are nilpotent and defined by

\be
P_1=\frac{\gamma_0-\gamma_5}2,~~~P_2=\frac{\gamma_0+\gamma_5}2\label{eq:3.6}
\ee
such that
\be
P_1^2=P_2^2=0,~~~[P_1.P_2]_+=-1
\ee

 After substitution in the constraint $\chi$ we find the condition on the states
\be
(-i\gamma_0 p^0\gamma_5 - M\gamma_5)\Psi(p^0)=0
\ee
Multiplying by $\gamma_5$
\be
(i\gamma_0 p^0-M)\Psi(p^0)=0
\ee
This shows that the model describes at the same time particles and antiparticles corresponding to the two eigenvalues 
$\pm 1$ of $i\gamma_0$.

Since there are only 3 operators, the lowest dimensional representation is in terms of  $2\times 2$matrices. For instance, we can take the following representation which satisfied the previous quantization conditions
\be
\hat\pi^5 =i\sqrt{\frac M2}\frac {(i\sigma_3-\sigma_2)}2,~~~ \hat\xi^5 =\sqrt{\frac 2 M}\frac {(i\sigma_3+\sigma_2)}2,~~~\tilde\xi_0=\frac i{\sqrt{2M}}\sigma_1\ee
where the $\sigma_i$'s are the Pauli matrices. Then the odd constrint $\chi$ becomes
\be
(-iM\sigma_2+p^0\sigma_1)\Psi(p^0)=0
\ee
Multiplying by $\sigma_1$
\be
(p^0+M\sigma_3)\Psi(p^0)=0
\ee
 Note that we have an ultralocal Dirac equation.

\section{The VSUSY Galilei  particle}

%Following the strategy of  ref. \cite{Barducci:2018wuj}, 
We start with a lagrangian (\ref{eq:4.0}) invariant  under the VSUSY algebra on the total space-time in $D+1$ dimensions \cite{Barducci:1976qu,Casalbuoni:2008iy}, in its euclidean version
\be
L=M\sqrt{(\dot x^\mu-i\xi^\mu\dot\xi^5)^2}-\beta\left(\dot c+\frac i2\xi_\mu\dot \xi^\mu\right)-\gamma\left(\dot c_5+\frac i2\xi^5\dot\xi^5\right)
\ee
Let us consider a $D$-contraction of the Galilei type for $r-t=0$. %(see Section \ref{sec:3.4}. 
According to our philosophy, 
the Galilei invariant lagrangian is obtained by using the previous lagrangian  in the euclidean subspace in $D$ dimensions, and then adding its variation under a boost  times a lagrangian multiplier. Since in this case the boost is operating only on the space-coordinates, $\delta\vec  x=\vec v x^0$ (see eq. (\ref{eq:3.72})), we get
\be
L=M\sqrt{(\dot{\vec x}-i\vec\xi\dot\xi^5)^2}-\beta\left(\dot c+\frac i2\vec\xi\cdot\dot{\vec\xi}\right)-\gamma\left(\dot c_5+\frac i2\xi^5\dot\xi^5\right)+\lambda \dot x^0\label{eq:2.1}
\ee
This lagrangian  is in fact VSUSY invariant under the following rigid transformations (see \cite{Casalbuoni:2008iy})
\be
\delta\vec x=i\vec\epsilon\,\xi^5,~~~\delta\vec\xi =\vec\epsilon,~~~ \delta\xi^5=\epsilon^5
\ee
\be
\delta c=-\frac i2 \vec\epsilon\cdot\vec\xi,~~~\delta c_5=-\frac i 2\epsilon^5\xi^5
\ee
The generators of these transformations satisfy the algebra
\be
[G_i,G_j]_+ =Z\delta_{ij},~~~[G_5,G_5]_+ =Z_5,~~~[G_i, G_5]_+ =-P_i
\ee
where $Z$ and $Z_5$ are two central charges. 
The variation of   (\ref{eq:2.1})  under Galilei bosts is given by (remember thgat $x^0$ does not transform under a Galilei boost)
\be
\delta L=\vec p\cdot\delta \dot{\vec  x} +\delta\lambda {\dot x}^0= \vec p\cdot\vec v{\dot x}^0 + +\delta\lambda {\dot x}^0
\ee\label{eq:5.6}
where $\vec p$ is the  canonical momentum
\be 
\vec p=M\frac{(\dot{\vec x}-i\vec\xi\dot\xi^5)}{\sqrt{(\dot{\vec x}-i\vec\xi\dot\xi^5)^2}}
\ee
Therefore the model is Galilei invariant if 
\be
\delta\vec\lambda=-\vec p\cdot\vec v 
\ee
The other  odd canonical momenta are
\be
\vec\pi=\frac{i\beta}2\vec\xi\label{eq:68}
\ee
\be \pi^5=\frac{i\gamma}2\xi^5 +i\vec p\cdot\vec\xi
\ee
From these equations the following constraints follow
\be
\phi=\vec p^{\,2}-M^2
\ee
\be
\chi=\pi^5-\frac{i\gamma}2\xi^5-i\vec p\cdot\vec \xi\label{eq:1.9}
\ee
\be
\vec\chi=\vec\pi-\frac{i\beta}2\vec\xi
\ee
The last constraint is  second class. We then define Dirac brackets in the usual way and, in particular, we find
\be
\{\xi^i,\xi^j\}^*=\frac i\beta\delta^{ij}\ee
We are now able to evaluate the Dirac brackets of the odd constraint $\chi$ with itself. We find
\be
\{\chi,\chi\}^*=-\frac i\beta(\vec p^{\,2}-\beta\gamma)\ee
Therefore, if we choose 
\be
\beta\gamma=M^2
\ee 
the constraints $\phi$ and $\chi$ are first class and we expect, as in the relativistic case, a local  kappa symmetry of the lagrangian.

This model represents the  spinning  pseudoclassical generalization of the galileian massless model studied 
in \cite{Sou,Duval:2009vt,Batlle:2017cfa}.  The model can be also obtained as the non-relativistic limit of a tachyonic spinning particle, see the appendix.

Although tbe two cases $r-t=0$ and $r-t=1$ correspond to two different algebras, it is easy  to check that  the lagrangian (\ref{eq:2.1}) is invariant  also under the algebra corresponding to $r-t=1$.
%{\color{red} is there any physical reason for this?}
\subsection{$\kappa$-invariance}

As we have seen, the model has two local symmetries, generated by the constraints $\phi$ and $\chi$, if the condition $\beta\gamma=M^2$ holds. We will satisfy this condition by  the choice $\beta=\gamma=M$.
The local symmetry generated by the odd constraint is a "$\kappa$-symmetry". The transformations generated by $\chi$ are given by
\be
\delta\vec x=i\kappa\vec\xi,~~~\delta\vec\xi=\kappa\frac{\vec p}M,~~~\delta\xi^5=-\kappa
\ee
%Then, by choosing
%\be
%\beta=\gamma=M
%\ee
we find
\bea
&\dd{\delta L=i\kappa(\vec p\cdot\dot{\vec\xi}-M\dot\xi^5)-\frac i 2\kappa\vec p\cdot\dot{\vec \xi}-\frac i2\vec\xi\cdot \frac d{d\tau}(\kappa\vec p)+\frac i2 M\kappa\dot\xi^5+\frac icM\xi^5\dot k}&\nn\\
&\dd{=\frac i2\kappa\vec p\cdot\dot{\vec\xi}+\frac i2\frac d{d\tau}(\kappa\vec p)\cdot\vec\xi+\frac i2M\xi^5\dot\kappa+\frac i2 M\kappa\dot\xi^5\kappa=\frac i2\frac d{d\tau}[\kappa(\vec p\cdot\vec\xi-M\xi^5)]}&
\eea
where we have used $\beta\gamma=M^2$. Therefore the lagrangian (\ref{eq:2.1}) is quasi-invariant under this $\kappa$-transformation.

%{\color{magenta} 
%The kappa symmetry allows to eliminate the variable $\xi^5$. 

%{\color{magenta} Concordo con Andrea, quanto segue non e' assolutamente chiaro. Nel caso si puo' anche omettere
%
%
%The VSUSY transformations and the kappa symmetry% in this gauge 
%are given by
%
%
% 
%\be
%\delta\vec x=0,~~~\delta\vec\xi =\vec\epsilon+\kappa\frac{\vec p}M,~~~ \delta\xi^5=\epsilon^5-\kappa=0
%\ee
%\be
%\delta c=-\frac i2 \vec\epsilon\cdot\vec\xi,~~~\delta c_5=-\frac i 2\epsilon^5\xi^5
%\ee
%The BPS solution are given the bosanic configuration
%${\vec p}=0$ with $\vec\epsilon=0$ and therefore preserves
%$\frac 14$ of supersymmetry.}

\subsection{Quantization}

To perform the canonical quantization
 we require 
 the following conditions for the operators corresponding to the Grassmann variables \cite{Casalbuoni:1975bj}
% Dynamical Grassmann variables are quantized via anticommutators white a value equal to the classical Poisson (or Dirac) brackets time the imagnary unit \cite{Casalbuoni:1975bj}:
% \be
% [.,.]_+ =i\{.,.\}
% \ee
%In the present case, by denoting with a hat the operators corresponding to the variables $\pi^5,\xi^5,\vec\xi$. 
%we have
\be
[\hat\pi^5,\hat\pi^5]_+=[\hat\xi^5,\hat\xi^5]_+=[\hat\xi^5,\hat\xi_i]_+=[\hat\xi_i,\hat\pi^5]_+=0,~~~[\hat\pi^5,\hat\xi^5]_+=-i,~~~[\hat\xi_i,\hat\xi_j]_+=-\frac 1 M\delta_{ij}\label{eq:3.2}
\ee
In this context we will suppose to be in a space-time of even dimensions $D+1=2m$. The $\gamma$-matrices satisfy 
\be
[\gamma_\mu,\gamma_\nu]_+ = 2g_{\mu\nu}
\ee

Then, we see that the anticommutators in eq. (\ref{eq:3.2}) are satisfied by 
\be
\hat\pi^5=i\sqrt{\frac M2}P_1,~~~\hat\xi^5=\sqrt{\frac 2 M}P_2,~~~\hat\xi_i=\frac i{\sqrt{2M}}\gamma_i
\ee
where $P_1$ and $P_2$ are defined in (\ref{eq:3.6}) .
Substituting these expressions inside the constraint (\ref{eq:1.9}) we find that the physical states must satisfy the equation
\be
(\vec p\cdot\vec\gamma-iM\gamma_5)\Psi(\vec p)=0
\ee
or, multiplying by $i\gamma_5$
\be
(i\vec p\cdot\vec\gamma\gamma_5-M)\Psi(\vec p)=0
\ee
Then, define
\be
\Gamma_i=i\gamma_i\gamma_5
\ee
which satisfy
\be
[\Gamma_i,\Gamma_j]_+=2\delta_{ij}
\ee
Therefore the $\Gamma_i$'s give a representation of the Clifford algebra in a space of $D$ dimensions equivalent to the one spanned by the $\gamma_i$'s. In this way we get
\be
(\vec p\cdot\vec\Gamma-M)\Psi(\vec p)=0
\ee
which is the Dirac equation in an euclidean space of  odd dimensions $D$.  After this equatione is satisfied, the physical states satisfy automatically the condition arising fron the even constraint $\phi$
\be
(\vec p^{\,2}-M^2) \Psi(\vec p)=0
\ee
Notice also that the lower dimensional representation of the Clifford algebra in  $D$ odd dimensions is through  $2^{(D-1)/2}\times 2^{(D-1)/2}$ matrices. For instance in 3-dimensions the lowest dimensional representation is by the $2\times 2$ Pauli matrices.

%{\color{magenta}  Are there models of two spinning particles like DGL? }

\section{Conclusions}

%{\color{blue}
In this paper we have introduced two models of  pseudoclassical \cite{Casalbuoni:1975bj} spinning particles respectively invariant under the Carroll and the Galilei groups with zero central charge. The construction of these models has been done through a generalization of the method introduced in ref. \cite{Barducci:2018wuj}. 
The method consists in starting from a space-time in $D+1$ dimensions and partitioning it in two parts, the first minkowskian and the second euclidean. Then,  in the  Carroll case an
 invariant model can be obtained by introducing a Minkowski invariant action in the first part of the space-time and in the second part a system of lagrange multipliers transforming in an appropriate way  under the euclidean group 
 such to confine the system to a region of the space-time.
  An analogous procedure can be followed in the Galilei case.

%This method is based on the generalized contractions of the Poincar\'e algebra, to Carroll and Galilei algebras. 
This procedure  allows to construct a class of
Carroll and Galilei invariant models, once given a relativistic action in configuration space without performing a Carroll or non-relativistic limit.
%represents an easy way of construvting Carroll and Galilei invariant models, once given the relativistic action. 
In the two  particular cases discussed here, we have started from an action 
 in configuration space invariant under the VSUSY algebra \cite{Barducci:1976qu} \cite{Casalbuoni:2008iy}. %Therefore 
%We have also presented the generalized contractions of the VSUSY algebra consistent with the contractions of the Poincar\'e subalgebra discussed in \cite{Barducci:2018wuj}.
%
The two models    turn out to be the  pseudoclassical spinning version of the Carroll particle described in ref. \cite{Bergshoeff:2014jla} and of the Galilei particle of refs. \cite{Sou,Duval:2009vt,Batlle:2017cfa}. 
%However, as explaine in the introduction, our descriprion is madse in the configuration space, contrarily to the scalar cases which have been formulated in phase space. 
We have also checked that our results can be reproduced  starting from the VSUSY invariant model and performing the standard Carroll and Galilei limiting procedure.
%We have also presented 
 The limiting procedure is consistent with
the generalized contractions of the VSUSY algebra consistent with the contractions of the Poincar\'e subalgebra discussed in \cite{Barducci:2018wuj}.

We have performed a detailed analysis of these models, showing that they preserve the same features of the VSUSY invariant model. Namely, the presence of two constraints, one even corresponding to the mass-shell condition and  an odd one, leading to the relevant analogous of the Dirac equation, after quantization. We have also shown that with a convenient choice of the parameters appearing in the lagrangians, the models both possess a $\kappa$-symmetry.

%}

\section*{Acknowledgments}

 J. G. has been supported in part
by Programa Nacional de F\'{\i}sica de Part\'{\i}culas (FPA) 2013-
46570-C2-1-P and Consolider Centro Nacional de F\'{\i}sica de
Partículas, Astropart\'{\i}culas y Nuclear (CPAN) and by the
Spanish government (MINECO/FEDER) under project
MDM-2014-0369 of ICCUB (Unidad de Excelencia
María de Maeztu). J. G. has also been supported by
CONICYT under grant PAI801620047 as a visiting professor
of the Universidad Austral de Chile.

% JG has been supported in part by MINECO FPA2016-76005-C2-1-P and Consolider CPAN and by the Spanish goverment (MINECO/FEDER) under 
% project MDM-2014-0369 of ICCUB (Unidad de Excelencia Mar\'\i a de Maeztu).
% JG has also been  been suported by  CONICYT under grant PAI801620047 as a
% visiting professor of the Universidad Austral de Chile.  

\section{Appendix A}

In this Appendix we will show that for the two models presented here, our strategy is equivalent to take the usual Galilei and Carroll limit from the relativistic phase space lagrangiam

The canonical lagrangian of a relativistic spinning tachion is given by
\be L^C=p_\mu(\dot x^\mu-{i}\xi^\mu\dot\xi^5)
-\beta\frac{i}2\xi_\mu\dot\xi^\mu-\gamma\frac{i}2\xi^5\dot\xi^5
-\frac{e}2(p^2-m^2),\label{LagC}\ee
or in symplectic form
\be L^S=p_\mu\dot x^\mu
-\beta\frac{i}2\xi_\mu\dot\xi^\mu-\pi_5\dot\xi^5
-\frac{e}2(p^2-m^2)
-\rho(\pi_5-\gamma\frac{i}2\xi^5-ip_\mu\xi^\mu),\label{LagC1}
\ee
where
\be
\chi_5=\pi_5-\gamma\frac{i}2\xi^5-ip_\mu\xi^\mu
\ee
The Poisson-Dirac brackets are
\be
\{p_\mu,x^\nu\}^*=-{\delta_\mu}^\nu,\qquad
\{\xi^\mu,\xi^\nu\}^*=\frac{i}\beta{\eta^{\mu\nu}},\qquad
\{\pi_5,\xi^5\}^*=-1 \label{DB}
\ee
\be
\{\chi_5,\chi_5\}^*=
i\gamma-\frac{i}{\beta}p_\mu p_\nu\eta^{\mu\nu}=-\frac{2i}{\beta}\frac{e}2(p^2-m^2)\;.
\ee
where $\beta=\gamma=m$.
 $\chi_5$ generates the local kappa variation.
  
\subsection{Non-relativistic (Galilei) limit}

We define the non-relativistic limit as
\be
x^0=\omega t,~~~ p_0=-\frac{E}{\omega},~~~ \xi^\mu=\xi^\mu,~~~
\xi^5=\xi^5, ~~~\pi_5=\pi_5
\ee
in the limit $\omega\rightarrow\infty$ we have the action
\be L^{NR}=-E \dot t+\vec p\dot{\vec x}
-\beta\frac{i}2\xi_\mu\dot\xi^\mu-\pi_5\dot\xi^5
-\frac{e}2({\vec p}^{\,2}-m^2)
-\rho(\pi_5-\gamma\frac{i}2\xi^5-i{\vec p}\vec{\xi}),\label{LagC2}
\ee
which in the reduced space is equivalent to the hamiltonian form of (\ref{eq:2.1}).
The structure of Poisson brackets is mantained for the non-relativisitic variables.
Also the first class character of  constraints is maintained. 

\subsection{Carroll limit}
The symplectic action of a relativistic particle is
\be L^S=p_\mu\dot x^\mu
-\beta\frac{i}2\xi_\mu\dot\xi^\mu-\pi_5\dot\xi^5
-\frac{e}2(p^2+m^2)
-\rho(\pi_5-\gamma\frac{i}2\xi^5-ip_\mu\xi^\mu),\label{LagCa1}
\ee
where $\beta=-m, \gamma=m$

We define the Carroll limit as
\be
x^0=\frac{1}{\omega} t,~~~ p_0=-{\omega}{E},~~~ \xi^\mu=\frac{1}{\sqrt{\omega}}\xi^\mu,~~~
\xi^5=\frac{1}{\sqrt{\omega}}\xi^5, ~~~\pi_5=\omega\pi_5,~~~ m=\omega M
\ee
in the limit $\omega\rightarrow\infty$ we have the action
\be L^{Carroll}=-E \dot t+\vec p\dot{\vec x}
-\beta\frac{i}2\xi_\mu\dot\xi^\mu-\pi_5\dot\xi^5
-\frac{\tilde e}2(E^2-M^2)
-\tilde\rho(\pi_5-\gamma\frac{i}2\xi^5+iE {\xi}^0),\label{LagCa2}
\ee
where $\tilde e=-\omega^2 e, \tilde\rho=\sqrt{\omega}\rho$,
.which in the reduced space is equivalent to the hamiltonian form of (\ref{eq:2.1}).
The Poisson-Dirac brackets and the character  of the constraints is maintained.

\end{document}